\documentclass[10pt]{iopart}
\usepackage{graphicx}
\usepackage{xurl}
\usepackage{amssymb}
\begin{document}
\title[]{$^{171}$Yb$^+$ optical clock with $2.2\times 10^{-18}$ systematic uncertainty and absolute frequency measurements}

\author{
A Tofful$^{1,2}$,
C F A Baynham$^1$\footnote{Present address:
Department of Physics, Blackett Laboratory, Imperial College London, London SW7~2AZ, United Kingdom.},
E A Curtis$^1$,
A O Parsons$^1$\footnote{Present address: Culham Science Centre, Abingdon OX14~3EB, United Kingdom.},
B I Robertson$^1$\footnote{Present address: ORCA Computing, LG, 30 Eastbourne Terrace, London W2~6LA, United Kingdom.},
M Schioppo$^1$,
J Tunesi$^1$,
H S Margolis$^1$,
R J Hendricks$^1$,
J Whale$^1$,
R C Thompson$^2$
and R M Godun$^1$}

\address{$^1$ Time and Frequency Department, National Physical Laboratory, Hampton Road,
Teddington TW11~0LW, UK.}
\address{$^2$ Department of Physics, Blackett Laboratory,
Imperial College London, London SW7~2AZ, UK}
\ead{alexandra.tofful@npl.co.uk}

\begin{abstract}
A full evaluation of the uncertainty budget for the ytterbium ion optical clock at the National Physical Laboratory (NPL) was performed on the electric octupole (E3) $^2\mathrm{S}_{1/2}\,\rightarrow\, ^2\mathrm{F}_{7/2}$ transition. The total systematic frequency shift was measured with a fractional standard systematic uncertainty of $2.2\times 10^{-18}$. Furthermore, the absolute frequency of the E3 transition of the $^{171}$Yb$^+$ ion was measured between 2019 and 2023 via a link to International Atomic Time (TAI) and against the local caesium fountain NPL-CsF2. The absolute frequencies were measured with fractional standard uncertainties between $3.7 \times 10^{-16}$ and $1.1 \times 10^{-15}$, and all were in agreement with the 2021 BIPM recommended frequency.
\end{abstract}
Keywords: optical clock, ytterbium ion, uncertainty budget, absolute frequency

\maketitle
\ioptwocol

\section{Introduction}
Optical atomic clocks have been demonstrated to reach fractional systematic frequency uncertainties below the $10^{-18}$ level \cite{Brewer2019}, outperforming caesium microwave clocks by two orders of magnitude. Because of the high level of achievable accuracy and stability of optical clocks, the uncertainty with which their absolute frequencies can be measured is limited to the $10^{-16}$ level by the uncertainty of caesium primary frequency standards. For this reason, the Consultative Committee for Time and Frequency (CCTF) has developed a roadmap towards redefining the second in terms of an optical frequency standard \cite{Dimarcq2024}.

The ytterbium ion, $^{171}$Yb$^+$ \cite{Huntemann2016, baynham2018}, is an excellent candidate for a primary optical frequency standard thanks to the availability of the strongly forbidden electric octupole (E3) transition $^2\mathrm{S}_{1/2}\,\rightarrow\, ^2\mathrm{F}_{7/2}$, which has an extremely narrow natural linewidth of the order of nHz \cite{Hosaka2005, Webster2002,Lange2021}. This transition is also relatively insensitive to perturbations from external electric and magnetic fields. Another feature of $^{171}$Yb$^+$ is the presence of a second favourable optical clock transition, the electric quadrupole (E2) transition $^2\mathrm{S}_{1/2}\,\rightarrow\, ^2\mathrm{D}_{3/2}$~\cite{schneider05}, with a narrow enough natural linewidth (3.1~Hz) to make it suitable as a frequency standard, which can be utilised to help determine systematic effects on the E3 transition. Furthermore, the E3 transition has a strong sensitivity to the variation of the fine structure constant, which makes $^{171}$Yb$^+$ ideal for testing fundamental physics \cite{Godun2014, Huntemann14, Filtzinger2023, Sherrill2023}.

The stability and accuracy of optical clocks at levels below 1 part in $10^{16}$ can only be evaluated via direct frequency ratio measurements between two optical clocks. However, it is also important to measure the absolute frequency of optical clock transitions with low enough uncertainty to show reproducibility and ensure continuity between the current and the future definition of the second. Absolute frequency measurements can either be performed with a direct link to a caesium fountain realising the SI second, or via a satellite link to International Atomic Time (TAI).

In this paper, section~\ref{sec:experiment} describes the experimental setup of the $^{171}$Yb$^+$ ion optical clock at the National Physical Laboratory (NPL). Section~\ref{sec:systematics} explains the methods for evaluating the known systematic frequency shifts experienced by the E3 transition in the ion. Finally, section~\ref{sec:results} reports full systematic uncertainty budgets for the E3 transition of the $^{171}$Yb$^+$ ion optical clock, together with the associated measurements of its absolute frequency both against a local caesium fountain and via a link to TAI.
\section{Experimental setup}\label{sec:experiment}
A detailed description of the $^{171}$Yb$^+$ ion experimental setup at NPL is presented in \cite{tofful2023}. This section summarises the key points.

A single $^{171}$Yb$^+$ ion is trapped in a radio frequency (RF) Paul end-cap trap \cite{Schrama1993} designed to minimise black-body radiation from the trap, phonon heating of the ion, ion micromotion caused by stray electric fields, and background gas collisions \cite{NisbetJones2016}. The trapping potential is generated by two cylindrical molybdenum end-cap electrodes to which a trapping frequency of $\Omega_{\mathrm{RF}}=2\pi\times 13.7$ MHz is applied through a helical resonator and conducted by a copper support structure in which the end caps are mounted. The end caps are enclosed within a pair of conical DC electrodes, which, together with an additional two compensation electrodes, provide the voltages necessary for the minimisation of ion micromotion. The trap is contained within an ultra-high-vacuum chamber operating at a pressure of $\sim~1\times10^{-11}$~mbar, maintained by a non-evaporable getter pump and a diode ion pump. Coils in a near-Helmholtz configuration are wound around the vacuum chamber, enabling magnetic field control at the ion.

\begin{figure}[t]
    \centering
    \includegraphics[width=0.9\linewidth]{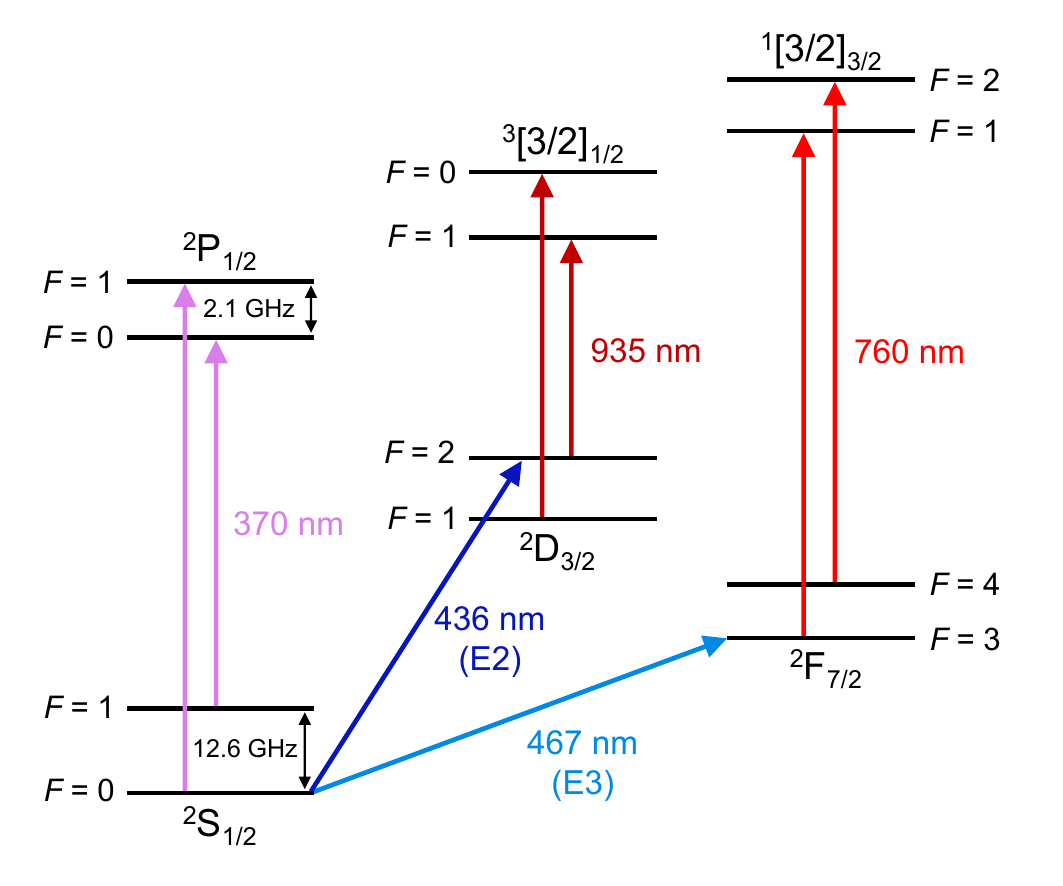}
    \caption{Atomic level structure of the $^{171}$Yb$^+$ ion.}
    \label{fig:yb}
\end{figure}

The ion is loaded into the trap by the following method: an atomic oven containing isotopically-enriched $^{171}$Yb is resistively heated, and then in a two-photon ionisation process, 399-nm and 370-nm light interacts with the collimated atomic beam at the centre of the trap to photoionise $^{171}$Yb atoms. The ion is Doppler cooled by three near-orthogonal 370-nm beams driving the $^2\mathrm{S}_{1/2}\,\rightarrow\,^2\mathrm{P}_{1/2}$ transition. To prevent the ion from being optically pumped into a non-resonant state, a magnetic field of $\sim 300$ \textmu T is applied for the destabilisation of these dark states \cite{Berkeland2002}, increasing the cooling rate. Moreover, one of the most likely routes for the ion to leave the cooling cycle is via spontaneous decay to the $^2\mathrm{D}_{3/2}$ metastable state, which is mitigated by repumping with 935-nm light. All the beams are combined by a series of dichroic mirrors and coupled by a parabolic collimator into an endlessly-single-mode polarisation-maintaining photonic crystal fibre (PCF) which delivers the light to the ion. The atomic level diagram of $^{171}$Yb$^+$ showing the relevant wavelengths for cooling, repumping, and probing the clock transition is displayed in figure~\ref{fig:yb}.

The E3 clock transition $^2\mathrm{S}_{1/2} (F=0) \rightarrow~^2\mathrm{F}_{7/2} (F=3)$ is probed by a 467-nm beam which is frequency-doubled from a 934-nm source. The frequency of the 934-nm laser is stabilised in two stages: first, it is locked to a temperature-controlled ultra-low-expansion glass cavity via a Pound-Drever-Hall lock \cite{PDH}, and secondly, the stability of a 1064-nm master laser locked to a 48.5-cm long ultrastable cavity ($<7\times10^{-17}$ fractional instability at 1 s) \cite{Schioppo2022} is transferred via an optical frequency comb to the 934-nm light using a transfer oscillator scheme \cite{Telle2002}. This high level of frequency stability enables the E3 transition to be coherently probed for over 500 ms.

Rabi spectroscopy is used to probe the clock transition. A single clock cycle is composed of six stages, starting from Doppler cooling. Next, the ion is optically pumped such that it is in the $^2\mathrm{S}_{1/2}\,(F=0)$ ground state of the clock transition. A pre-probe delay of 25 ms is applied to extinguish all laser light at the ion, and to ensure sufficient decay of the magnetic field which was aiding the ion cooling. The E3 transition is then probed with a Rabi $\pi$-pulse by switching on an acousto-optical modulator (AOM) in the 467-nm beam, which is intensity-stabilised, with a typical probe duration ranging from 320 to 650 ms. During the probe, the ion experiences a bias magnetic field chosen to align its quantisation axis for maximal excitation probability. The state of the ion is then read out by turning on the 370-nm and 935-nm beams and observing the fluorescence from the ion cycling between the ground state and the upper cooling $^2\mathrm{P}_{1/2}$ state, captured by a photo-multiplier tube (PMT). A lack of fluorescence in the readout stage indicates shelving in the $^2\mathrm{F}_{7/2} (F=3)$ state due to the successful excitation of the clock transition. Finally, a 760-nm beam is turned on to ensure that the ion is repumped out of the upper clock state back into the cooling cycle.

In order to lock the 467-nm light to the clock transition, a servo control method is employed \cite{tofful2023}. Rabi pulse probes are alternated in a Thue-Morse sequence \cite{Schat2015} on the high and low frequency sides of the atomic resonance at the half maximum points. At the end of each servo cycle, typically made up of 4 probes on each side of the transition, this information is fed back via the aforementioned AOM, whose driving RF signal is responsible for applying the appropriate frequency in each probe. Any asymmetry in the excitation fraction gives information about the magnitude and direction of the 467-nm frequency offset, such that a new probe frequency can be estimated. The AOM frequency and the beat between the 934-nm light and a frequency comb are recorded by a high-resolution phase and frequency counter, which is referenced to a 10-MHz signal coming from a free-running hydrogen maser that can be traced back to the local time scale UTC(NPL).

The optical clock experiment is run by a computer-controlled FPGA-based framework \cite{artiq_specs} which enables a high level of automation of the experiment, and facilitates the continuous monitoring of many parameters. When the ion fluorescence drops and remains below a set threshold, indicating that the ion has gone dark, a recovery protocol is automatically launched to bring the clock back into operation. The algorithm steps through a series of checks based on the monitored metrics to determine how to recover ion fluorescence, up to and including automated reloading of an ion into the trap. 

Potential causes for a dark ion are molecule formation \cite{Hoang2020} or a strong collision that excites the ion into a higher motional state. To simultaneously account for both possible issues, the 399-nm beam is turned on to attempt to dissociate the molecule at the same time as a 370-nm beam 220 MHz far-red-detuned from the $^2\mathrm{S}_{1/2}\,(F=1)\,\rightarrow\,^2\mathrm{P}_{1/2}\,(F=0)$ transition is turned on for additional cooling. If the ion fluorescence is observed to rise above the threshold, the lock to the clock transition is then automatically resumed. However, if the fluorescence does not return after 15 minutes, the ion is considered lost and the ion-loading experimental routine is automatically triggered. Upon loading a new ion, the lock is resumed. The implementation of this algorithm had a very positive impact on the uptime of the clock operation, as shown in section~\ref{sec:uptime}.

\section{Systematic frequency shifts}\label{sec:systematics}
This section will present all the known systematic frequency shifts of the E3 transition of the $^{171}$Yb$^+$ ion, and how they were measured and evaluated at NPL. The results will be presented in section~\ref{sec:results} as part of the uncertainty budget tables.
\subsection{Micromotion-related shifts}
In a trapping potential driven at a frequency $\Omega_{\mathrm{RF}}$, the ion is not perfectly motionless. A mean displacement from the ion trap's RF node gives rise to excess micromotion (EMM), which can be minimised by an appropriate choice of voltages in the four DC electrodes~\cite{NisbetJones2016}. Furthermore, the temperature-induced secular motion of the ion in the harmonic potential also gives rise to intrinsic micromotion (IMM). The secular and micromotion effects contribute to both a second-order Doppler shift and an RF Stark shift.

For an ion of mass $m$ and temperature $T$ trapped in an RF potential driven at frequency $\Omega_{\mathrm{RF}}$, the fractional second-order Doppler shift caused by the ion's motion is given by \cite{Keller2015}
\begin{equation}\label{eq:doppler}
    \left\langle\frac{\Delta\nu_{\mathrm{Dop.}}}{\nu}\right\rangle=-\left(\frac{e}{2mc\Omega_{\mathrm{RF}}}\mathbf{E}_{\mathrm{RF}}\right)^2-\frac{3k_\mathrm{B}T}{mc^2}\;.
\end{equation}
The first term arises from motion driven by the residual RF field at the mean position of the ion, $\mathbf{E}_{\mathrm{RF}}$, and the second term is the contribution from the thermal kinetic energy of the ion with RF confinement in all three directions.

The electric-field-induced micromotion leads to an RF Stark shift proportional to $\langle\mathbf{E}^2\rangle$ \cite{Dube2013}:
\begin{equation}\label{eq:rf_stark}
    \left\langle\frac{\Delta\nu_{\mathrm{RF}}}{\nu}\right\rangle=-\frac{\Delta\alpha}{2h\nu}\left(\frac{\mathbf{E}_{\mathrm{RF}}^2}{2}+\frac{m\Omega_{\mathrm{RF}}^2}{e^2}3k_\mathrm{B}T\right)\;,
\end{equation}
where $\Delta\alpha=\alpha_{\mathrm{excited}}-\alpha_{\mathrm{ground}}$ is the differential scalar polarisability of the clock transition. In this equation, the first and second terms represent the contributions from EMM and IMM respectively.

The ion micromotion is measured with the technique of RF-photon correlation \cite{Keller2015}, which consists of correlating the phase of the RF trap drive with the EMM-induced fluorescence modulation in the three near-orthogonal cooling beams. Daily minimisation of the micromotion is carried out with iterative corrections to the DC electrode voltages, typically achieving fluorescence modulation depths below 2.5~\%. The measured modulation depths are used to determine the ion's driven motion and hence $\mathbf{E}_{\mathrm{RF}}$ in the equations above.

The effective temperature of the ion was determined by measuring the ratio $R$ of the sideband-to-carrier excitation intensity of the E2 clock transition. Since the transition is probed in the horizontal direction which is perpendicular to the axial trap dimension, this measurement was performed on the two radial sidebands with frequencies $\omega_{x}$ and $\omega_{y}$, and their temperature contributions are combined according to the following equation \cite{Dube2013}:
\begin{equation}\label{eq:ion_temp}
    T=\frac{2mc^2}{k_B}\sum_{i=x,y,z}\left(\frac{\omega_{i}}{\omega_0}\right)^2R_i\;,
\end{equation}
where $m$ is the mass of the ion and $\omega_0$ is the angular frequency of the probe laser. $R_z=0$ due to the fact that the probing beam is in the radial direction. Equation~\ref{eq:ion_temp} is valid under the assumptions that the energy in the three secular modes is the same, and that the three modes are orthogonal.

\begin{figure}
    \centering
    \includegraphics[width=\linewidth]{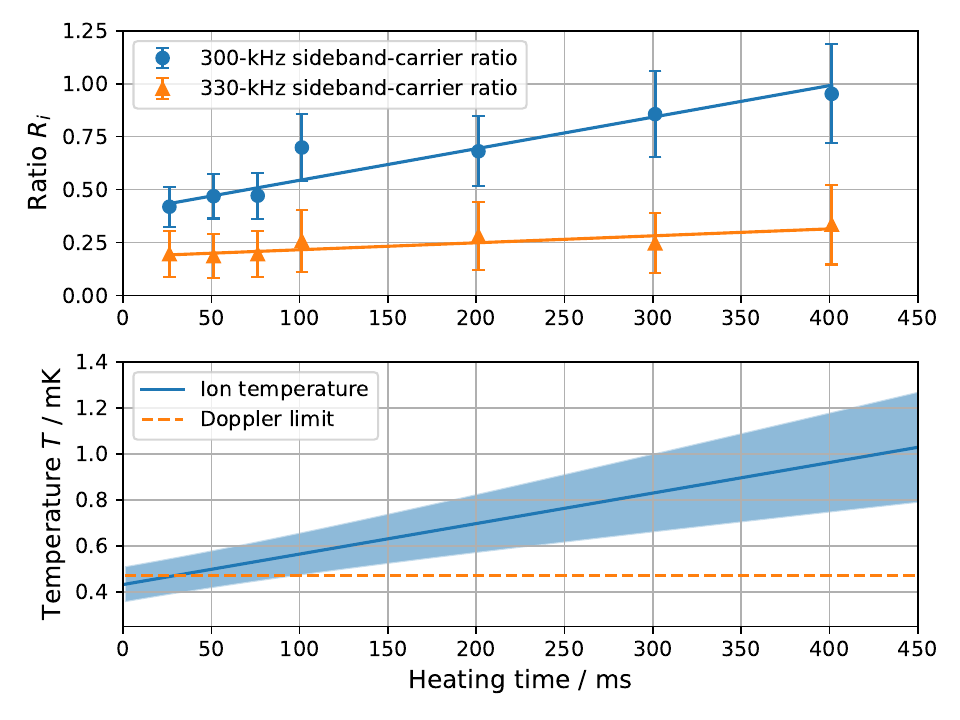}
    \caption{Top: sideband-carrier intensity ratios $R_x$ and $R_y$ measured at different ion heating times. Due to the probe beam direction being at 32$^{\circ}$ from $y$, it is expected for the ratio contributions sampled with this beam to be different for the two radial modes. Bottom: ion heating rate calculated using equation~\ref{eq:ion_temp} based on the ratios $R_i$ measured at different heating times.}
    \label{fig:sidebands}
\end{figure}

The estimated ion temperature is shown in figure \ref{fig:sidebands} as a function of time after active cooling is stopped (heating time). The individual sideband-carrier ratio contributions sampled on the two secular modes are also shown in the figure. The ion temperature immediately after cooling is found to be $0.43 \pm 0.08$~mK, which is consistent with the Doppler cooling limit of $0.47$~mK. Finally, the ion heating rate was measured to be $1.3 \pm 0.5$~mK/s. It is assumed that the three secular modes have the same heating rate. The mean temperature of the ion during a clock probe is calculated at time $t = t_{\mathrm{pre-probe\;delay}} + t_{\mathrm{probe}}/2$, and is equal to $0.73 \pm 0.14$~mK for a 400-ms probe. In equations~\ref{eq:doppler} and~\ref{eq:rf_stark}, the Doppler and RF Stark shifts are therefore dominated by the temperature-dependent term.
\subsection{Quadratic Zeeman shift from static fields}\label{sec:static_zeeman}
During an E3 clock probe, a 3-\textmu T bias magnetic field is applied in the designated $y$ direction to align the ion's quantisation axis for maximal excitation. Since the clock transition is driven between two $m_F=0$ states, it is insensitive to linear Zeeman shifts, and therefore only quadratic Zeeman shifts need to be characterised.

The background magnetic field in the laboratory is re-evaluated daily by performing spectroscopy scans of the E2 transition's $\Delta m_F=1$ and $\Delta m_F=2$ Zeeman components in six magnetic field directions in order to minimise the uncertainty in the magnitude and direction of the applied bias field. The E2 transition is exploited because the available laser power enables shorter probe times leading to faster measurements. It is found that the measured magnetic field magnitude $B_{\mathrm{DC}}$ varies by less than 30 nT day by day. The quadratic DC Zeeman shift is evaluated as:
\begin{equation}\label{eq:zeeman}
    \Delta\nu_{\mathrm{Z,DC}}=k_{\mathrm{E3}}B^2_{\mathrm{DC}}\;,
\end{equation}
where $k_{E3}=-2.08 \pm 0.01$ mHz/\textmu T$^2$ \cite{Godun2014}.

\subsection{Quadratic Zeeman shift from oscillating fields}
When evaluating quadratic Zeeman shifts, it is also important to account for the effect of oscillating magnetic fields, both parallel and orthogonal to the ion's quantisation axis. For AC fields that vary slower than the Larmor precession frequency ($\sim 50$ kHz), the ion's orientation will follow the magnetic field direction, so the magnetic field is always parallel to the ion's quantisation axis. For AC magnetic fields that vary faster than the Larmor precession frequency, it is important to distinguish between parallel and perpendicular components as they will have different quadratic Zeeman coefficients, $k$.
From the measurement of the linear Zeeman shift described in section~\ref{sec:static_zeeman}, the width of the Zeeman components (assumed to be broadened primarily by 50-Hz mains noise) is used to determine the low-frequency quadratic AC Zeeman shift contribution in the direction parallel to the quantisation axis:
\begin{equation}
    \Delta\nu_{\mathrm{Z,AC\parallel}}=k_{\mathrm{E3}}\langle B^2_{\mathrm{AC}}\rangle\;.
\end{equation}
where $\langle B^2_{\mathrm{AC}}\rangle$ is the time average of the square of the AC magnetic field.

Evaluating the shift from parallel and perpendicular components at higher frequencies needs more attention. In our system, the most likely cause of rapidly oscillating magnetic fields is from currents driven at the trapping frequency of $\Omega_{\mathrm{RF}}=2\pi \times 13.7$ MHz in the electrodes near the ion.

The magnitude of trap-induced AC magnetic fields was assessed following the method described in \cite{gan2018}, by driving the microwave transition between the $F=0$ and $F=1$ levels of the $^2$S$_{1/2}$ state and creating an Autler-Townes splitting \cite{autler1955}. When the linear Zeeman shift in the $F=1$ level is on resonance with the trap drive frequency, AC magnetic fields at the trap frequency can couple the $m_F=0$ state with the $m_F=\pm 1$ states. This leads to an Autler-Townes splitting $\Omega$ given by \cite{gan2018}:
\begin{equation}
    \Omega = \frac{g_F\mu_B B_{\perp}}{\hbar\sqrt{2}},
\end{equation}
where $\hbar$ is the reduced Planck constant, $g_F$ is the Land\'e g-factor, $\mu_B$ is the Bohr magneton and $B_{\perp}$ is the amplitude of the AC magnetic field perpendicular to the bias field direction during probing. The size of the splitting depends both on the RF trap drive amplitude and on the asymmetry of the trap geometry.

\begin{figure}
    \centering
    \includegraphics[width=1.1\linewidth]{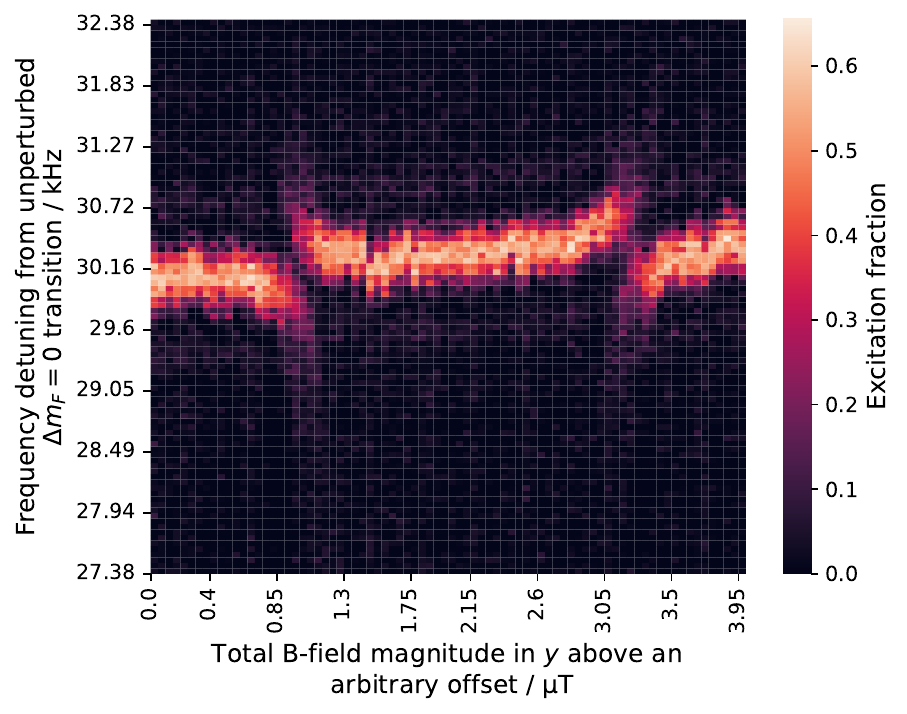}
    \caption{Measured Autler-Townes splittings as a function of magnetic field around the microwave transition $|F=0, m_F=0\rangle$ to $|F=1, m_F=0\rangle$. The two separate features arise when $B_{\perp}$ at the trapping frequency $\Omega_{\mathrm{RF}}=2\pi\times 13.7$ MHz resonantly couples the Zeeman states $m_F=0$ to $m_F=\pm 1$, respectively, in the $F=1$ level.}
    \label{fig:heatmap}
\end{figure}

In order to measure the magnitude of the Autler-Townes splitting, the microwave transition $^2\mathrm{S}_{1/2}(F=0 \rightarrow F=1)$ was probed with a 12.6-GHz microwave source whose output was guided by a microwave horn. A set of permanent magnets was positioned to create a large magnetic field ($\sim 1$ mT) in the $y$ direction, which, in addition to a variable magnetic field produced from a current-carrying coil wound around the trap itself, induced a linear Zeeman shift to match the $13.7$ MHz trap drive. A series of Rabi spectroscopy scans was carried out on the microwave carrier while making small variations to the magnitude of the applied magnetic field. This led to the observation of an Autler-Townes splitting of $\Omega=2\pi \times (570 \pm 140)$ Hz corresponding to $B_{\perp}=58 \pm 14$ nT.

Figure~\ref{fig:heatmap} shows two Autler-Townes splittings occurring at different magnetic fields approximately 2.2~\textmu T apart. The two features arise from the fact that there are unequal energy gaps between the three $m_F$ states in $|F=1\rangle$, since the $m_F=0$ state experiences a DC quadratic Zeeman shift while the $m_F=\pm1$ states do not. For this reason, the energy gaps $|1,1\rangle-|1,0\rangle$ and $|1,0\rangle-|1,-1\rangle$ are resonant with $\Omega_{\mathrm{RF}}$ at two different bias magnetic field magnitudes. 

Because this measurement only revealed the magnitude of the magnetic field perpendicular to the $y$ direction, a second measurement was performed with the strong magnets repositioned at an angle of $37^{\circ}\pm5^{\circ}$ around the $z$ axis in the $x$-$y$ plane. This made it possible to set a bound on the magnitude of the AC magnetic field component parallel to the $y$ direction, $B_{\parallel}$. For the E3 transition probed with a bias magnetic field in the $y$ direction, the fractional trap-induced AC Zeeman shift and its uncertainty are both $\lesssim 1 \times 10^{-20}$. Further details on their evaluation are found in \cite{curtis2024}.

\subsection{Electric quadrupole shift}\label{sec:quadshift}
The electric quadrupole shift is the result of the interaction of the electric quadrupole moment of the ion with an electric field gradient. The quadrupole moment $\Theta(L,J)$ of the upper state $|L,J\rangle$ of the clock transition is aligned with the quantisation axis of the ion, whose direction is set by the bias magnetic field $\mathbf{B}$ at the ion. Moreover, an electric field gradient $\mathbf{\nabla E}$ is produced at the centre of the trapping region by the voltages in the four DC electrodes used for micromotion minimisation.

Assuming a cylindrically symmetric trap potential, the quadrupole shift has the following form:
\begin{equation}
    \Delta \nu_{\mathrm{Q}} = \frac{\Delta \nu_0}{2}\,(3\cos^2\rho-1)\;,
\end{equation}
where $\rho$ is the angle between the ion's quantisation axis and the direction of the electric field gradient at the ion, and $\Delta\nu_0$ is related to the magnitude of the shift, which can be expressed as
\begin{equation}
    \Delta\nu_0=|\nabla \mathbf{E}|\, k(J,I,F,m_F)\,\Theta(L,J)\;.
\end{equation}
For the E3 transition, $\Theta(F,7/2)=-0.0297 ea_0^2$ (where $e$ is the elementary charge and $a_0$ the Bohr radius) and $k=5/7$, while for the E2 transition, $\Theta(D,3/2)=~1.95ea_0^2$ and $k=1$ \cite{Lange2020}.

Since the E3 transition is probed in a single magnetic field direction $\textbf{B}_y$, the quadrupole shift is nonzero and must be directly evaluated. The main difficulty lies in the fact that the electric field gradient at the location of the ion cannot be directly measured, and must therefore be extracted from the observed quadrupole shift in several different magnetic field directions. For this measurement, the E2 transition is employed for its higher sensitivity to the quadrupole shift due to its larger quadrupole moment. The E2 transition is probed in five magnetic field directions, chosen to extensively cover the vector space, alternated every 30 minutes. This measurement is interleaved with a second measurement probing the E2 transition in only one of the five fields, which acts as a stable reference to cancel out the frequency drift of the probing laser.

With these data, it is possible to calculate a quadrupole shift for each of the five magnetic field directions. These measured quadrupole shifts, which are typically up to 1 Hz, are used as input to a fitting algorithm that determines the direction and magnitude of the electric field gradient, such that the quadrupole shift and uncertainty can be calculated for any magnetic field direction, in this case specifically $\textbf{B}_y$. The shift and uncertainty on the E2 transition can then be scaled down by a factor of $(-10.9 \pm 0.2) \times 10^{-3}$ to a shift and uncertainty on the E3 transition using the $\Theta(L,J)$ and $k$ factors stated previously.

A combination of DC electrode voltages that minimises the ion micromotion was chosen such that the resulting electric field gradient would lead to a relatively small quadrupole shift uncertainty, and at the same time have a small component in the $y$-direction, such that any drifts or fluctuations in the bias magnetic field $\textbf{B}_y$ would have a minimal effect on the shift. It was verified experimentally that magnetic field drifts and fluctuations throughout one month of clock operation had a negligible contribution to the quadrupole shift uncertainty in our chosen electric field gradient.

With a given electrode voltage configuration, the uncertainty on evaluating the quadrupole shift is currently statistics-limited. Therefore, using this measurement technique, a lower uncertainty can be achieved by running the quadrupole shift measurement for a longer duration. Currently, the measurements are typically acquired over a duration of $>5$ days, leading to fractional quadrupole shift uncertainties $< 2.4\times10^{-18}$. This quadrupole shift measurement is performed once before and once after an extended measurement of the E3 transition, and it was verified experimentally that the two results are consistent within their uncertainties at 1-month timescales.

\subsection{AC Stark shifts}
The E3 transition experiences an AC Stark shift arising from laser light which the ion is exposed to \cite{Itano}. This includes several contributions that are treated separately, namely the intensity-stabilised 467-nm light during the probe time, the power overshoot of the 467-nm clock excitation light at the start of the probe as the power stabilisation servo pulls in, and the leakage light through the closed shutters from all other beams during the probe.

The probe-induced AC Stark shift on the E3 transition can typically be up to tens of hertz due to the relatively high 467-nm intensity (up to 500 \textmu W focused to a spot size of 19 \textmu m) used for probing the transition. This is particularly problematic when targeting a measurement accuracy at the mHz level. For this reason, two interleaved servos locking the laser frequency to the E3 transition are run at two different power levels $P_1$ and $P_2$. These two different powers produce linearly-related AC Stark shifts, resulting in resonant-excitation frequencies $f_1$ and $f_2$. The AC-Stark-free frequency $f_0$ is therefore extrapolated after a pair of high and low-power probes, using the formula:
\begin{equation}\label{eq:extrapolated}
    f_0=f_2-\frac{R}{R-1}(f_2-f_1)\;,
\end{equation}
where $R=P_2/P_1$ and $P_1<P_2$. The two power levels are continuously monitored by an out-of-loop photodiode after the lasers have passed through the ion trapping region in order to determine the effective power ratio $R_{\mathrm{eff}}$, which can slightly fluctuate or drift over time due to imperfect power servo electronics. This monitoring enables a small residual AC Stark shift correction to be applied to $f_0$, whose uncertainty is given by:
\begin{equation}
    \sigma_\nu=\frac{\Delta f}{(R_{\mathrm{eff}}-1)^2}\sigma_R\;,
\end{equation}
where $\Delta f$ is the average frequency difference $f_2-f_1$, and the uncertainty $\sigma_R$ is the flicker floor of the overlapping Allan deviation of the monitored power ratio.
For the clock measurement conducted in 2023, presented in section~\ref{sec:results}, the probe durations used for the high and low-power servos were 320 ms and 480 ms respectively, and were chosen such that they would each produce a $\pi$-pulse at their selected optical powers with a power ratio of 3:1. The effective power ratio $R_{\mathrm{eff}}$ was measured to be $3.0260$ with an uncertainty $\sigma_R = 0.0002$ and the average frequency difference $\Delta f$ between the high and low-power servos was approximately 7 Hz.

The overshoot in the 467-nm power when the clock AOM turns on at the beginning of the pulse leads to an additional AC Stark shift. Due to the limited bandwidth of the integrator, there is a short period of time (tens to hundreds of microseconds) between the AOM switch-on and the moment the beam power is brought down to its requested power level, and this causes the clock transition frequency to shift temporarily. A small overshoot is necessary to ensure the fast engagement of the power servo. The coupling between the power overshoot and frequency shift is modelled as described in \cite{baynham}, and found in our system to lead to a fractional shift $< 1\times 10^{-19}$.

Finally, to determine the effect of the shutter-leakage light from the other beams on the clock transition frequency, a single-photon counting module (SPCM) was set up at the output of the final fibre before the ion to measure the photon counts from each individual input beam with their shutters closed. For each measurement, all other beams are additionally blocked completely, to ensure that only the shutter-leakage light from the beam in question is detected. The photon-count measurements are then scaled by the response of the SPCM to the particular wavelength.

\subsection{Black-body radiation shift}
The black-body radiation (BBR) Stark shift is caused by the interaction of the ion with an oscillating electric field $E$ originating from the room-temperature environmental radiation. The BBR shift at a temperature $T$ can be expressed as
\begin{equation}
    \Delta\nu_{\mathrm{BBR}}=-\frac{\Delta\alpha(1+\eta)}{2h}\langle E^2 \rangle\;,
\end{equation}
where $\Delta\alpha$ is the differential scalar polarisability, $\eta$ is a dynamic correction that accounts for the spectral distribution of the black-body radiation, and $\langle E^2 \rangle$ is the mean squared electric field~\cite{itano82}:
\begin{equation}
    \langle E^2 \rangle=(831.9\;\mathrm{V/m})^2\left(\frac{T}{300\;\mathrm{ K}}\right)^4\;.
\end{equation}
The temperature of the ion's environment during clock operation is carefully monitored using two calibrated Pt100 sensors placed in different locations on the exterior of the vacuum chamber. The effective temperature that the ion experiences in the trap has been well characterised \cite{NisbetJones2016}, and Pt100 readings recorded at 30-s intervals allow this shift to be cancelled dynamically in post-processing with a temperature uncertainty $\sim 0.1$ K.
The uncertainty on the BBR shift is currently limited by the uncertainty on $\Delta\alpha$. For the E3 transition, whose differential scalar polarisability is particularly small \cite{Huntemann2016}, the BBR shift uncertainty in our laboratory at a room temperature of 20.0 $^{\circ}$C is at the $1\times10^{-18}$ level. 

\subsection{Phase chirp shift}\label{sec:chirp}
The 467-nm probe light passes through free space, as well as through two fibres and one control AOM before reaching the ion. Phase noise will be accumulated along the entire optical path. Most of this noise will be random fluctuations, which may cause a slight frequency broadening of the light but do not lead to a frequency offset over long timescales. However, some effects such as AOM switching and shutter vibrations can lead to a phase chirp that is synchronised with the probing of the ion and may not average to zero over multiple probes. 

In order to measure the effect of the phase chirp, the 467-nm beam is split by a 50:50 beam splitter such that half of the power goes through the AOM and the other half does not, acting as the reference beam. These two beams are then recombined in free space and their beat signal is detected on a photodiode and mixed with a local oscillator. Asymmetric phase excursions were observed to occur in the first 22 ms of a clock pulse, and were found to be dominated by the vibrations from the shutters on the same breadboard opening and closing during the clock sequence. Without the shutters, it was found that the AOM phase chirp shift was negligible in comparison.

The corresponding frequency shift of the clock transition, $\Delta\nu_{\mathrm{chirp}}$, was calculated using the method outlined in \cite{Falke2012}. This method is based on using a sensitivity function to determine how phase excursions in the probe light lead to changes in excitation probability $p$. When probing the clock transition at $\nu_0\pm\nu_{\mathrm{FWHM}}/2$, the frequency shift of the transition, induced by a difference $\Delta p$ in excitation on the two sides of the full width half maximum (FWHM), is \cite{Falke2012}:
\begin{equation}
    \Delta\nu_{\mathrm{chirp}}=\frac{1}{2}\frac{\Delta p}{p_{\mathrm{max}}}\nu_{\mathrm{FWHM}}\;,
\end{equation}
where $p_{\mathrm{max}}$ is the peak excitation probability. The sensitivity function at the start of a 500-ms long Rabi pulse greatly suppresses the change in excitation probability, leading to an offset in the clock transition from a single probe at the low $10^{-19}$ level. Due to the sign of the asymmetric phase excursion changing over time, this offset was seen to average out over many clock cycles, and therefore the shift is set to zero and the uncertainty to this nominal offset.

\subsection{Servo offset}
A software servo is used to lock the frequency of the clock laser to the atomic transition. An error signal is derived from the imbalance of excitation probabilities when the atom is probed on the high and low frequency sides of the atomic resonance. The error signal is then used to calculate a correction to the clock laser frequency for each servo cycle of the clock. Since the clock laser is pre-stabilised to an optical cavity that drifts in frequency, there is the possibility of introducing a frequency offset in the clock output if the servo updates lag behind the clock laser's drifting frequency.

To minimise any servo offset, several precautions are taken: (i) the linear drift of the optical cavity is largely subtracted from the clock laser frequency before it probes the ion, leaving only residual drift that is below 1.5 mHz/s; (ii) the probing sequence on the ion is not a straight alternation between the high (H) and low (L) frequency sides of the transition, but instead probes HLLH LHHL in order to reduce the effects of 1st and 2nd order drifts; (iii) the servo algorithm has integral and double-integral gain terms to reduce offsets, (iv) after a lock is started, the first 15 minutes of data are discarded to avoid using data from the servo pull-in period (informed by the Allan deviation of the frequency data).

The servo offset and uncertainty in our datasets are calculated using a Monte Carlo simulation of hundreds of finite locks to the clock transition, mimicking the probing sequence and timings, with representative values for the cavity drift and servo gain parameters. Because the sign of the shift depends on the direction of the residual cavity drift, which could change throughout an extended measurement, the shift is set to zero. The uncertainty takes into account the steepest experienced cavity drift, and is calculated as the standard deviation of the mean frequency offsets in the simulations.
\subsection{Collisional shift}
Given that the vacuum in the chamber is not perfect, there is a possibility that a molecule from the background gas could collide with the trapped ion, temporarily shifting the clock transition frequency. Based on the Langevin collision model \cite{eichelberger2003}, it was estimated that these collisions occur at a rate of 1.5 h$^{-1}$ in our vacuum chamber with a pressure of $\sim 1\times 10^{-11}$ mbar \cite{baynham}. A conservative estimate is made by setting the uncertainty to the largest possible frequency shift caused by background collisions \cite{rosenband2008}, $6\times10^{-19}$ fractionally, and setting the actual shift to zero in the uncertainty budget. This estimate is in excellent agreement with the upper bound for the collisional shift proposed by \cite{vutha17} using a non-perturbative analysis method. 

\subsection{Relativistic redshift}
In order to compare the frequencies of two clocks in different gravity potentials, and also for referencing the clock frequency to the SI second, the relativistic redshift experienced by the ion must be evaluated to a high level of accuracy.
A combination of geodetic levelling and GNSS methods \cite{Denker2018} was used to determine the geopotential relative to the conventional equipotential $W_0=62\,636\,856.0$ $\mathrm{m}^2 /\mathrm{s}^2$~\cite{groten2000} at reference markers in a number of atomic clock laboratories, opening up the possibility to compare clocks in different parts of the world. With the NPL $^{171}$Yb$^+$ ion located at a height of $1.029 \pm 0.001$ m above a reference marker measured to be at a geopotential of $96.58\pm 0.22$~m$^2$/s$^2$ \cite{Riedel2020} and in a local gravitational acceleration of 9.8118 m/s$^2$, the fractional relativistic redshift on the E3 transition is equal to $(1186.9\pm 2.5)\times10^{-18}$.
\section{Measurements and results}\label{sec:results}
Most recently, the absolute frequency of the E3 transition was evaluated over the month of March 2023 by measuring the frequency of the clock laser with an optical frequency comb referenced to the 10-MHz signal of a hydrogen maser denomimated HM6. The maser acts as a link to both the local caesium fountain primary frequency standard and TAI. This section will present the results of the measurements, starting with the evaluated systematic uncertainty budget for the measurement period and the uptime of the optical clock operation. Next, the different corrections and uncertainty contributions of the links to both TAI and the caesium fountain will be presented. Finally, the results from two other extended optical clock measurements from 2019 and 2022 will be summarised.
\subsection{Systematic uncertainty budget in 2023}
The systematic uncertainty budget associated with the March 2023 measurement of the E3 transition is reported in table~\ref{tab:2023_budget}, with a total relative standard uncertainty of $2.2\times10^{-18}$. The leading uncertainty is the electric quadrupole shift, whose accuracy is currently limited by the statistical uncertainty on the E2 frequency measurement in the five magnetic fields, as discussed in section~\ref{sec:quadshift}. The uncertainty of $1.5\times10^{-18}$ is the average of a 5-day-long and a 7-day-long measurement respectively before and after the E3 measurement campaign.
\begin{table}[b]
\caption[Uncertainty budget of the E3 clock transition frequency of $^{171}$Yb$^+$ during the 2023 measurement campaign.]{Uncertainty budget of the E3 clock transition frequency of $^{171}$Yb$^+$ during the 2023 measurement campaign (excluding the relativistic redshift). The fractional uncertainties are reported as 1-sigma confidence intervals. $^*$These shifts were dynamically corrected, and this is taken into account when summing towards the total shift.}
\label{tab:2023_budget}
\begin{indented}
\item[]\begin{tabular}{@{}lll}
\br
\textbf{Systematic effect}   & \textbf{$\Delta\nu/\nu_0\;[10^{-18}]$} & \textbf{$\sigma/\nu_0\;[10^{-18}]$} \\
\mr
Electric quadrupole & 22.8 & 1.5 \\
Black-body radiation & $-66.6^*$ & 1.2 \\
Quadratic Zeeman (DC) & $-$29.2 & 0.6 \\
Background gas collisions & 0 & 0.6 \\
Phase chirp & 0 & 0.5 \\
AC Stark - probe beam & 6758$^*$ & 0.4 \\
Second-order Doppler & $-1.3^*$ & 0.3 \\
Trapping RF Stark & $-0.30^*$ & 0.06 \\
Servo offset & 0 & 0.05 \\
Quadratic Zeeman (AC) & $-0.14$ & 0.05 \\
Trap-induced AC Zeeman & $<0.01$ & $<0.01$ \\
AC Stark - overshoot & 0.08 & $<0.01$ \\
AC Stark - leakage light & $<0.01$ & $<0.01$ \\ 
\mr
\textbf{Total} & $\mathbf{-6.5}$ & $\mathbf{2.2}$ \\
\br
\end{tabular}
\end{indented}
\end{table}

The second highest systematic uncertainty comes from the BBR shift, whose limiting factor is the uncertainty of the current best estimate of the differential scalar polarisability coefficient \cite{Huntemann2016}. The shift is dynamically corrected in post-processing based on temperature measurements collected at 30-second intervals. The uncertainty is calculated based on the temperature measurements filtered by the uptime of the optical clock. While the RF Stark and second-order Doppler shifts were nearly constant across the whole one-month dataset, they were dynamically corrected in post-processing to account for a single weekend during which there was a noticeable deviation due to the excess micromotion amplitude being higher than usual (an average fluorescence modulation depth of 4.5~\% in the three beam directions, exceeding the calibration limit of 2.5~\%). Finally, it must be noted that while most of the probe-induced AC Stark shift is eliminated in real time by means of the dual-servo lock configuration, the residual shift is also dynamically corrected in post-processing.

\subsection{Clock Uptime}\label{sec:uptime}
Optical clock operation with high uptime is crucial for several reasons. When the uncertainty of the frequency measurement is statistics-limited, longer operation means that a lower instability can be reached. This is particularly important for the measurement of optical frequency ratios between different clocks, where their uptime overlap needs to be maximised for the same reason. Moreover, it is important for the absolute frequency measurement via TAI to achieve a high uptime over the month-long evaluation period, since gaps in the data can lead to additional uncertainties, as will be explained in section~\ref{sec:tai}.
\begin{figure}[b]
    \centering
    \includegraphics[width=\linewidth]{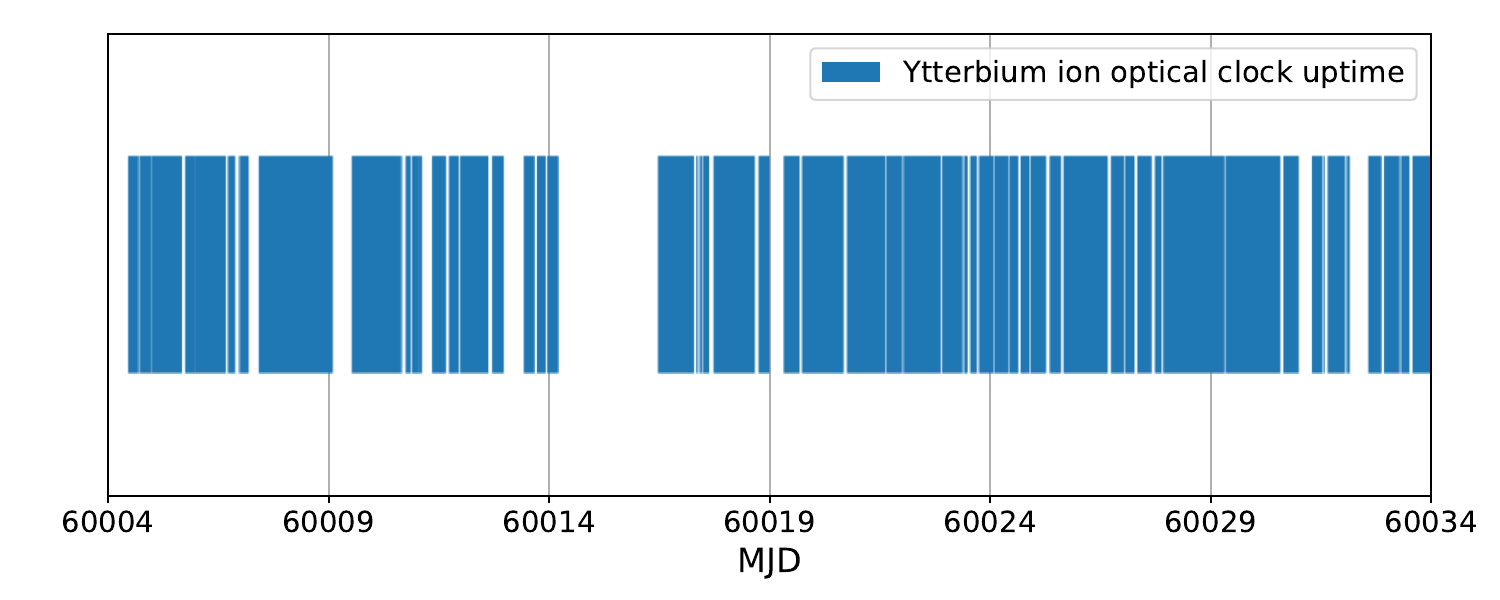}
    \caption{Uptime of the $^{171}$Yb$^+$ ion optical clock (excluding frequency comb downtime) over the month of March 2023.}
    \label{fig:uptime}
\end{figure}

The $^{171}$Yb$^+$ ion optical clock uptime (excluding frequency comb downtime) during the month of March 2023 totalled 76~\%, and is shown in figure~\ref{fig:uptime}. Optical clock data is flagged as valid when the symmetric exponentially weighted moving average (SEWMA) of the excitation fraction of the E3 transition is between 10~\% and 70~\%. Additionally, once the SEWMA crosses either threshold, the previous 5 minutes of data are invalidated. Finally, any time the lock to the clock transition is started or resumed, the first 15 minutes are invalidated. 

The main sources of downtime were due to experiment interruptions for routine systematic shift minimisation, and the unexpected failure of a power supply essential to the experiment. Furthermore, excluding the instance of the hardware failure, the ion and lock recovery algorithm was demonstrated to have a 100~\% success rate every time it ran during this campaign. However, due to certain software processes at the time out of our control slowing down the computer, the recovery process would occasionally fail to launch, causing an unnecessary loss of uptime.
In most cases, failure modes could be automatically overcome thanks to the implementation of the recovery protocol described in section~\ref{sec:experiment}, which enabled clock operation to resume within a maximum of 30 minutes. The highest uptime achieved over a 10-day period was 92.5~\% between MJD 60020 and 60030. These encouraging results represent a big step forward towards the goal of unattended optical clock operation, which is one of the ancillary conditions for the optical redefinition of the second \cite{Dimarcq2024}.

\subsection{Absolute frequency measurement via TAI}\label{sec:tai}
TAI is a virtual time scale referenced to the SI second realised on the conventional equipotential $W_0=62\,636\,856.0$ $\mathrm{m}^2 /\mathrm{s}^2$. It is the result of the continuous comparison between hundreds of atomic clocks and hydrogen masers across the world, whose monthly weighted average realises the free-running time scale \'Echelle Atomique Libre (EAL). EAL is compared with the operational primary and secondary frequency standards to provide a frequency correction for TAI to conform with the definition of the SI second \cite{Petit2015,Panfilo2019}. TAI is computed monthly by the International Bureau of Weights and Measures (BIPM), which publishes the offsets between Coordinated Universal Time (UTC) and each local realisation of the time scale in the Circular T bulletin at 5-day intervals.

The following equation describes how in principle the absolute frequency of the $^{171}$Yb$^+$ E3 transition is referenced to the SI second via TAI:
\begin{equation}\label{eq:tai}
    \frac{f_{\mathrm{Yb}^+}}{f_{\mathrm{SI}}}=\frac{f_{\mathrm{Yb}^+}}{f_{\mathrm{UTC(NPL)}}}\times \frac{f_{\mathrm{UTC(NPL)}}}{f_{\mathrm{TAI}}}\times \frac{f_{\mathrm{TAI}}}{f_{\mathrm{SI}}}\;.
\end{equation}
At NPL, the hydrogen maser HM6 that provides the reference for the frequency comb is also responsible for the generation of the local time scale UTC(NPL), and it is continuously compared to TAI via satellite links. The problem with equation~\ref{eq:tai} is that it is only valid when the optical clock is operated with 100~\% uptime during the one-month TAI evaluation period. Because this is typically not the case, additional corrections need to be applied to extrapolate the frequency from a discontinuous optical clock measurement to the continuous TAI computation period.
A more general form of equation~\ref{eq:tai} can be expressed as:
\begin{equation*}
    \frac{f_{\mathrm{Yb}^+}}{f_{\mathrm{SI}}}=\frac{f_{\mathrm{Yb}^+;\Delta t_1}}{f_{\mathrm{UTC(NPL)};\Delta t_1}}\times \frac{f_{\mathrm{UTC(NPL)};\Delta t_1}}{f_{\mathrm{UTC(NPL)};\Delta t_2}}
\end{equation*}
\begin{equation}\label{eq:tai-full}
    \qquad \, \times \; \frac{f_{\mathrm{UTC(NPL)};\Delta t_2}}{f_{\mathrm{TAI};\Delta t_2}} \times \frac{f_{\mathrm{TAI};\Delta t_2}}{f_{\mathrm{TAI};\Delta t_3}}\times \frac{f_{\mathrm{TAI};\Delta t_3}}{f_{\mathrm{SI}}}\;,
\end{equation}
where $\Delta t_1$ represents the optical clock uptime on a 1-second grid, $\Delta t_2$ the 5-day reporting periods that the clock uptime overlaps with, and $\Delta t_3$ the 1-month TAI evaluation period. 
The method used for this absolute frequency measurement analysis has been previously presented in \cite{baynham2018, Hobson2020, Hachisu2015}, and a more detailed description of the method can be found in chapter 5 of \cite{tofful2023}.

Table~\ref{tab:2023_tai} shows the individual contribution from each ratio from equation~\ref{eq:tai-full} to the absolute frequency of the E3 transition calculated over the period $\Delta t_2=$ MJD 60004 -- 60034. The total optical clock uptime $\Delta t_1$ (allowing for frequency comb downtime) during this period was 66~\%.
For this particular measurement, the optical clock frequency was referenced to the unsteered output of HM6, whereas UTC(NPL) was generated by the steered output of HM6, and therefore there is an additional step in the chain to relate $f_{\mathrm{maser}}$ to $f_{\mathrm{UTC(NPL)}}$. Moreover, while the maser frequency is 10 MHz, $f_{\mathrm{maser}}$ and $f_{\mathrm{UTC(NPL)}}$ are considered to be scaled to 1 Hz for clarity.

\begin{figure}[b]
    \centering
    \includegraphics[width=\linewidth]{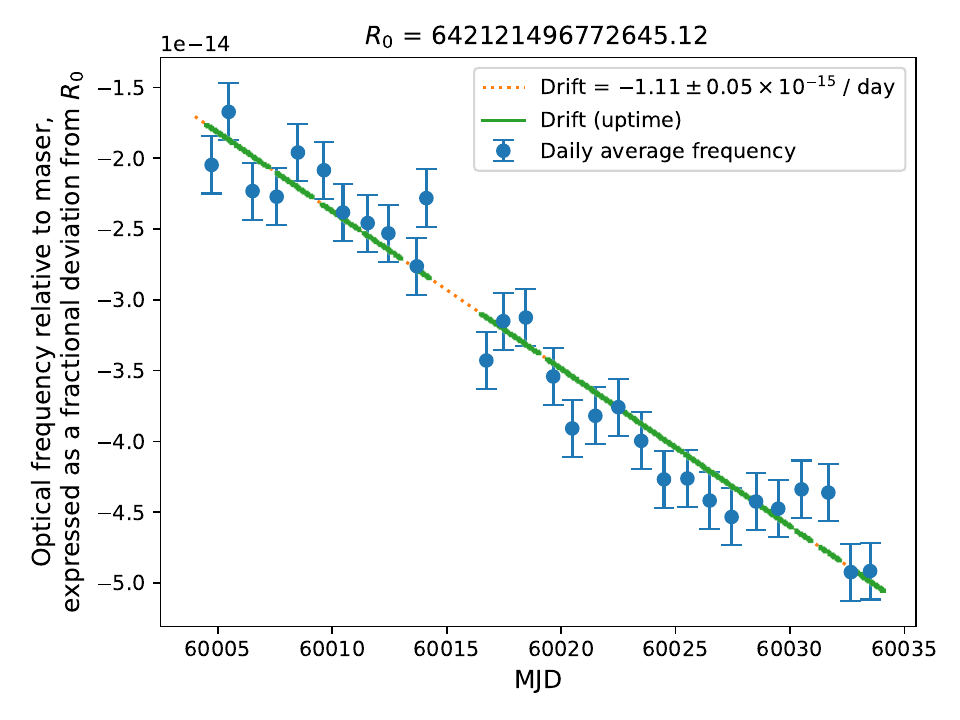}
    \caption{Frequency of the $^{171}$Yb$^+$ optical clock relative to the hydrogen maser measured in March 2023 offset from $R_0=642\,121\,496\,772\,645.12$.}
    \label{fig:maser-drift}
\end{figure}

As discussed in \cite{baynham2018, tofful2023}, a deterministic correction must be applied when extrapolating the optical clock frequency during periods of downtime based on the linear drift of the maser, which is in the opposite direction to the drift of the optical clock frequency relative to the maser, as shown in figure~\ref{fig:maser-drift}.

\begin{figure}
    \centering
    \includegraphics[width=\linewidth]{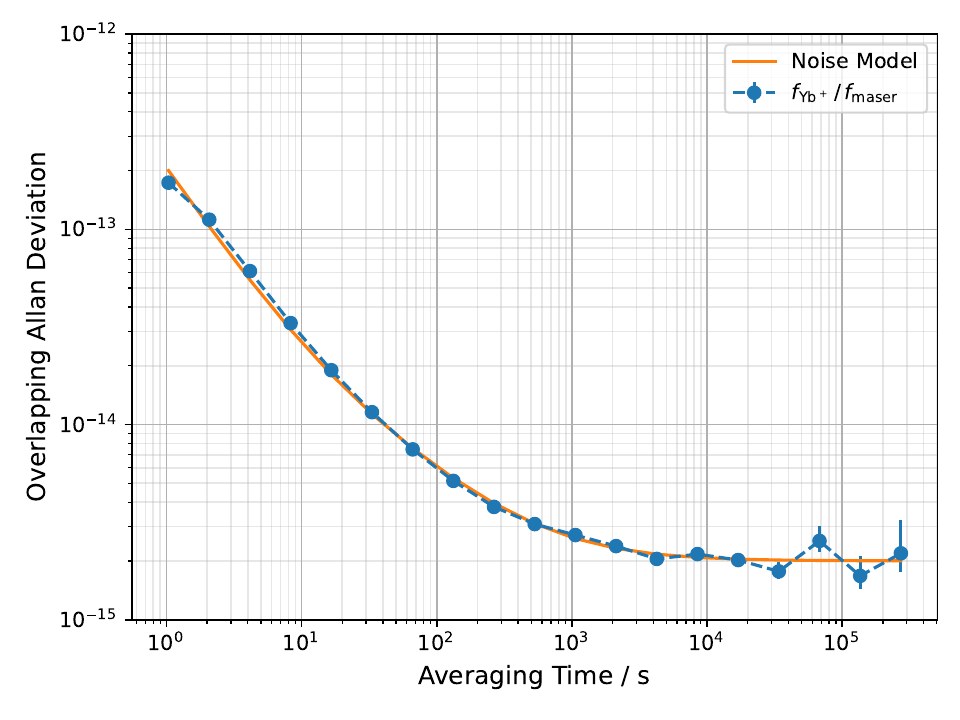}
    \caption{Measured instability and noise model of the hydrogen maser HM6 used in the 2023 absolute frequency measurement based on a combination of white phase noise, white frequency noise and flicker
frequency noise.}
    \label{fig:maser-noise}
\end{figure}
\begin{figure}
    \centering
    \includegraphics[width=\linewidth]{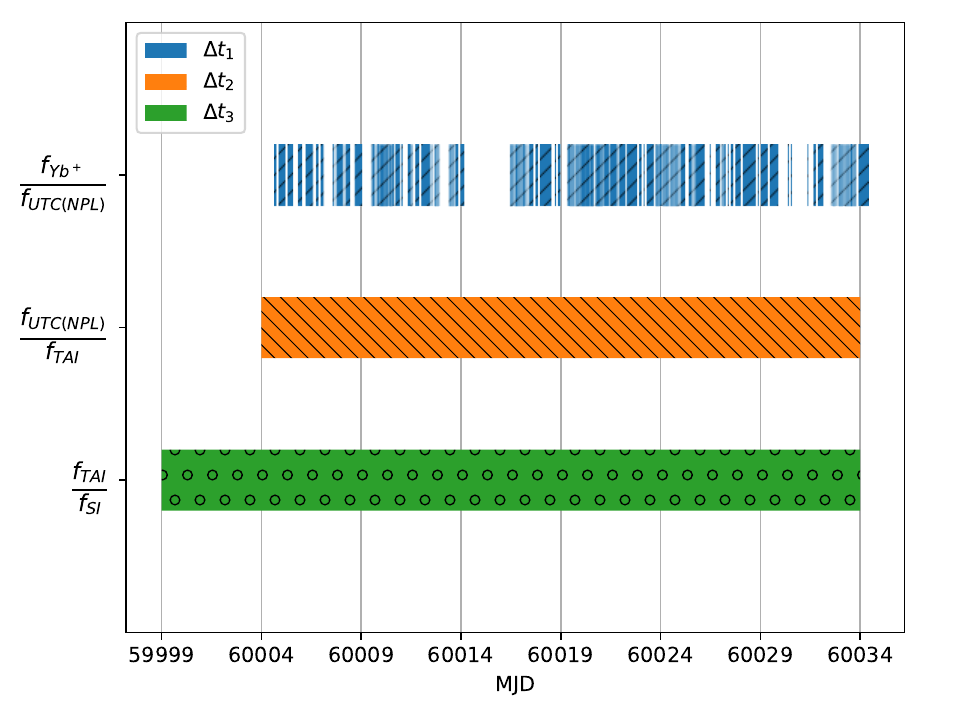}
    \caption[Time intervals for the computation of the corresponding frequency ratios during the 2023 measurement.]{Time intervals for the computation of the corresponding frequency ratios. $\Delta t_1$ is the combined uptime of the $^{171}$Yb$^+$ ion optical clock and frequency comb during the March 2023 measurement campaign. $\Delta t_2$ consists of six 5-day TAI reporting periods, while $\Delta t_3$ covers the 1-month TAI computation period.}
    \label{fig:time-intervals-2023}
\end{figure}

\begin{table*}[t]
\centering
\caption{Values and uncertainties associated with each ratio that contributes to the absolute frequency calculation of the E3 transition of $^{171}$Yb$^+$ over the period MJD 60004 -- 60034 (March 2023) according to equation~\ref{eq:tai-full}. $R_0=f_0/(1\;\mathrm{Hz})=642\,121\,496\,772\,645.12$, where $f_0$ is the recommended value of the frequency of the E3 transition approved by the International Committee for Weights and Measures (CIPM) in 2021.}
\label{tab:2023_tai}
\begin{center}
\resizebox{0.75\textwidth}{!}{
\begin{tabular}{@{}lllll}
\br
\textbf{Ratio}   & \textbf{Contribution} & $\mathbf{r_0}$   & $\mathbf{\left(\frac{r}{r_0}-1\right)\;[10^{-18}]}$ & $\mathbf{u\left(\frac{r}{r_0}-1\right)\;[10^{-18}]}$ \\
\mr
$\frac{f_{\mathrm{Yb}^+;\Delta t_1}}{f_{\mathrm{maser};\Delta t_1}}$ & Ratio at comb & $R_0$ & $-$34\,442         & 58                \\
        & Yb$^+$ statistics   &     & --            & 2                  \\
        & Yb$^+$ systematics   &     & $7$         & 2                \\
        & Relativistic redshift &  & $-$1187        & 2                  \\ \hline
$\frac{f_{\mathrm{maser};\Delta t_1}}{f_{\mathrm{maser};\Delta t_2}}$ & Maser noise      & 1     & --            & 278                \\
        & Maser drift       &       & 419           & 18                 \\ \hline
$\frac{f_{\mathrm{maser};\Delta t_2}}{f_{\mathrm{UTC(NPL)};\Delta t_2}}$ & Maser to UTC(NPL) & $1$      & 35\,257          & 29                \\ \hline
$\frac{f_{\mathrm{UTC(NPL)};\Delta t_2}}{f_{\mathrm{TAI};\Delta t_2}}$ & UTC(NPL) to TAI  & 1 & $-$77         & 196                \\ \hline
$\frac{f_{\mathrm{TAI};\Delta t_2}}{f_{\mathrm{TAI};\Delta t_3}}$ & EAL extrapolation  & 1     & --            & 8                 \\ \hline
$\frac{f_{\mathrm{TAI};\Delta t_3}}{f_{\mathrm{SI};\Delta t_3}}$ & TAI to SI second     & 1           & 100          & 120                \\
\mr
$\mathbf{\frac{f_{\mathrm{Yb}^+}}{f_{\mathrm{SI}}}}$   & \textbf{Total}        & $\mathbf{R_0}$          & \textbf{77}         & \textbf{367}                \\
\br
\end{tabular}}
\end{center}
\end{table*}
The dominant uncertainty contribution comes from the maser noise (shown in figure~\ref{fig:maser-noise}), which needs to be extrapolated from a discontinuous measurement to a continuous time period. The link between UTC(NPL) and TAI is another large source of uncertainty which increases with decreasing overlap between the optical clock measurement and the TAI evaluation period, which in this case was $\Delta t_3=$ MJD 59999 -- 60034. Furthermore, since the fractional frequency offset between TAI and the SI second ($d$-value) is reported in Circular T only for the full evaluation period, an additional uncertainty is introduced for the extrapolation of EAL in the 5-day interval(s) with no overlap. Figure~\ref{fig:time-intervals-2023} shows the three relevant time intervals for the computation of the absolute frequency and their relative overlaps. The absolute frequency of the E3 transition is measured to be $642\,121\,496\,772\,645.17(24)$ Hz, which is in excellent agreement with the 2021 BIPM recommended frequency of $642\,121\,496\,772\,645.12(12)$ Hz \cite{recommended_E3}.

\subsection{Absolute frequency via Cs fountain in 2023}
The local caesium fountain NPL-CsF2 \cite{Szymaniec2016}, which was operating during March 2023, enabled the absolute frequency measurement of the E3 transition to be performed with direct traceability to the SI second. The optical clock was compared against the caesium fountain via a hydrogen maser, whose 10 MHz signal was continuously compared with the local oscillator of the caesium fountain. Table~\ref{tab:2023_cs} shows a breakdown of the uncertainty contributions. In order to calculate the statistical uncertainty on the ratio, the overlapping uptime of $^{171}$Yb$^+$ E3 and CsF2 is considered. The overlapping Allan deviation for this particular dataset shows a hint of flattening out at long timescales, and therefore this level is considered as a conservative estimate of the statistical uncertainty. Over the period MJD~60004~--~60035, the absolute frequency is calculated to be $642\,121\,496\,772\,645.33(29)$ Hz.
\begin{table}[t]
\caption{Shift and uncertainty contributions to the absolute frequency of the $^{171}$Yb$^+$ E3 transition relative to NPL-CsF2 over the period MJD 60004 -- 60035 (March 2023), where $R_0'=f_{0_{\mathrm{Yb}^+}}/f_{0_{\mathrm{Cs}}}=69\,851.758\,760\,554\,065$.}
\label{tab:2023_cs}
\begin{indented}
\item[]\begin{tabular}{@{}llll}
\br
\textbf{Contribution}   & $\mathbf{r_0}$  & {\begin{tabular}{@{}l@{}}$\mathbf{(r/r_0-1)}$ \\ $\mathbf{[10^{-18}]}$\end{tabular}} & {\begin{tabular}{@{}l@{}}$\mathbf{u(r/r_0-1)}$ \\ $\mathbf{[10^{-18}}]$\end{tabular}} \\
\mr
Yb$^+$/CsF2 ratio & $R_0$ & 252  & 396 \\
Yb$^+$ systematics & 1  & $-7$ & 2 \\
CsF2 systematics    & 1  & 0 & 209 \\
Diff. relativistic redshift  & 1  & 84        & 2\\
Maser RF distribution  & 1  & -- & 58\\
\mr
\textbf{Total} & $\mathbf{R_0'}$  & $\mathbf{329}$ & $\mathbf{452}$ \\
\br
\end{tabular}
\end{indented}
\end{table}
\subsection{Absolute frequency results from 2019 and 2022}
The absolute frequency of the E3 transition was also measured in July 2019 against TAI and in March 2022 against both TAI and CsF2. These results, together with the associated systematic  uncertainty budgets, will be presented below, with any differences in the experimental and analysis methods compared to the 2023 measurement being described.
\begin{table}[b]
\caption[Uncertainty budget of the E3 clock transition frequency of $^{171}$Yb$^+$ during the 2022 measurement campaign.]{Uncertainty budget of the E3 clock transition frequency of $^{171}$Yb$^+$ during the 2022 measurement campaign (excluding the relativistic redshift). The fractional uncertainties are reported as 1-sigma confidence intervals. $^*$These shifts were dynamically corrected, and this is taken into account when summing towards the total shift.}
\label{tab:2022_budget}
\begin{indented}
\item[]\begin{tabular}{@{}lll}
\br
\textbf{Systematic effect}   & \textbf{$\Delta\nu/\nu_0\;[10^{-18}]$} & \textbf{$\sigma/\nu_0\;[10^{-18}]$} \\
\mr
Electric quadrupole & 22.3 & 2.8 \\
Black-body radiation & $-66.4^*$ & 1.2 \\
Background gas collisions & 0 & 0.6 \\
Quadratic Zeeman (DC) & $-28.8^*$ & 0.6 \\ 
Second-order Doppler & $-$1.5 & 0.4 \\
AC Stark - probe beam & 6448$^*$ & 0.3 \\ 
Phase chirp & 0 & 0.2 \\
Quadratic Zeeman (AC) & $-0.3$ & 0.2 \\
Trapping RF Stark & $-0.34$ & $0.07$ \\
AC Stark - overshoot & 0.07 & 0.07 \\
Servo offset & 0 & 0.06 \\
Trap-induced AC Zeeman & $<0.01$ & $<0.01$ \\
AC Stark - leakage light & $<0.01$ & $<0.01$ \\ 
\mr
\textbf{Total} & $\mathbf{20.3}$ & $\mathbf{3.2}$ \\
\br
\end{tabular}
\end{indented}
\end{table}

The systematic uncertainty budget for the March 2022 measurement is shown in table~\ref{tab:2022_budget}. The systematic shift measurement and analysis method is the same for both 2022 and 2023, and the main differences in the resulting values come from slightly different experimental parameters, such as electric and magnetic fields, probe powers and probe durations. Specifically, a probe duration of 650 ms was chosen for both the high and low-power servos, which was respectively slightly higher and lower than the probe duration required for achieving a $\pi$-pulse at the chosen optical powers.
\begin{table*}[ht!]
\centering
\caption{Values and uncertainties associated to each ratio that contributes to the absolute frequency calculation of the E3 transition of $^{171}$Yb$^+$ over the period MJD 59649 -- 59669 (March 2022) according to equation~\ref{eq:tai-full}, where $R_0=642\,121\,496\,772\,645.12$.}
\label{tab:2022_tai}
\begin{center}
\resizebox{0.75\textwidth}{!}{
\begin{tabular}{@{}lllll}
\br
\textbf{Ratio}   & \textbf{Contribution} & $\mathbf{r_0}$   & $\mathbf{\left(\frac{r}{r_0}-1\right)\;[10^{-18}]}$ & $\mathbf{u\left(\frac{r}{r_0}-1\right)\;[10^{-18}]}$ \\
\mr
$\frac{f_{\mathrm{Yb}^+;\Delta t_1}}{f_{\mathrm{maser};\Delta t_1}}$ & Ratio at comb & $R_0$ & 59\,477         & 51                \\
        & Yb$^+$ statistical   &     & --            & 8                  \\
        & Yb$^+$ systematics   &     & $-$20         & 3                \\
        & Relativistic redshift &  & $-$1187        & 2                  \\ \hline
$\frac{f_{\mathrm{maser};\Delta t_1}}{f_{\mathrm{maser};\Delta t_2}}$ & Maser noise      & 1     & --            & 899                \\
        & Maser drift       &       & $-$1152           & 102                 \\ \hline
$\frac{f_{\mathrm{maser};\Delta t_2}}{f_{\mathrm{UTC(NPL)};\Delta t_2}}$ & Maser to UTC(NPL) & 1      & $-$56\,654          & 44                \\ \hline
$\frac{f_{\mathrm{UTC(NPL)};\Delta t_2}}{f_{\mathrm{TAI};\Delta t_2}}$ & UTC(NPL) to TAI  & 1 & $-$116         & 282                \\ \hline
$\frac{f_{\mathrm{TAI};\Delta t_2}}{f_{\mathrm{TAI};\Delta t_3}}$ & EAL extrapolation  & 1     & --            & 25                 \\ \hline
$\frac{f_{\mathrm{TAI};\Delta t_3}}{f_{\mathrm{SI};\Delta t_3}}$ & TAI to SI second     & 1           & 160          & 120                \\
\mr
$\mathbf{\frac{f_{\mathrm{Yb}^+}}{f_{\mathrm{SI}}}}$   & \textbf{Total}        & $\mathbf{R_0}$           & \textbf{508}         & \textbf{958}                \\
\br
\end{tabular}}
\end{center}
\end{table*}

Once again, the systematic uncertainty is dominated by the quadrupole shift, and it is higher than in 2023 due to a shorter measurement dataset on the E2 transition, leading to a higher instability in the statistics-limited measurement. Furthermore, the quadratic DC Zeeman shift was dynamically corrected in post-processing to address the fact that an insufficient time was left for the eddy currents in the coils around the trap to dissipate before a clock probe. As a result, a slightly different magnetic field was experienced by the ion, depending on the pre-probe delay set, but taken into account in the correction protocol.

The absolute frequency of the E3 transition via TAI was calculated to be $642\,121\,496\,772\,645.45(62)$~Hz over MJD 59649 -- 59669, representing four 5-day TAI reporting periods, during which the optical clock had a 43~\% uptime. The individual uncertainty contributions are shown in table~\ref{tab:2022_tai}.
For this measurement, the uncertainty on the absolute frequency is strongly dominated by the noise of the maser, which in this particular time period exhibited diurnal temperature fluctuations. This required an artificial inflation of the maser noise flicker floor in the noise model used when extrapolating maser data over different time periods. It is worth noting that in this absolute frequency measurement, the frequency comb measuring the optical clock frequency was referenced to the unsteered signal of HM6, while UTC(NPL) was generated by the steered output of a different maser, HM4. The absolute frequency was also measured against NPL-CsF2 over the period MJD 59647 -- 59671 to be $642\,121\,496\,772\,645.10(42)$ Hz, and the individual uncertainty contributions can be found in table~\ref{tab:2022_cs}. For this measurement, the statistical uncertainty was calculated by fitting the white frequency noise component of the Allan deviation and extrapolating it to the total length of the dataset. This is justified by ample evidence from long-term caesium fountain data around this period showing that its statistical uncertainty averages down as $1/\sqrt{\tau}$.

\begin{table}[t]
\caption{Shift and uncertainty contributions to the absolute frequency of the $^{171}$Yb$^+$ E3 transition relative to NPL-CsF2 over the period MJD 59647 -- 59671 (March 2022), where $R_0'=f_{0_{\mathrm{Yb}^+}}/f_{0_{\mathrm{Cs}}}=69\,851.758\,760\,554\,065$.}
\label{tab:2022_cs}
\begin{indented}
\item[]\begin{tabular}{@{}llll}
\br
\textbf{Contribution}   & $\mathbf{r_0}$  & {\begin{tabular}{@{}l@{}}$\mathbf{(r/r_0-1)}$ \\ $\mathbf{[10^{-18}]}$\end{tabular}} & {\begin{tabular}{@{}l@{}}$\mathbf{u(r/r_0-1)}$ \\ $\mathbf{[10^{-18}}]$\end{tabular}} \\
\mr
Yb$^+$/CsF2 ratio & $R_0$ & $-140$  & 211 \\
Yb$^+$ systematics & 1 & 20 & 3 \\
CsF systematics    & 1 & 0 & 620 \\
Diff. relativistic redshift  & 1 & 84        & 2\\
Maser RF distribution  & 1 & -- & 51\\ 
\mr
\textbf{Total} & $\mathbf{R_0'}$ & $\mathbf{-36}$ & $\mathbf{657}$ \\
\br
\end{tabular}
\end{indented}
\end{table}
\begin{table}[t]
\caption{Uncertainty budget of the E3 clock transition frequency of $^{171}$Yb$^+$ during the 2019 measurement campaign (excluding the relativistic redshift). The fractional uncertainties are reported as 1-sigma confidence intervals. $^*$This shift was dynamically corrected, and this is taken into account when summing towards the total shift.}
\label{tab:2019_budget}
\begin{indented}
\item[]\begin{tabular}{@{}lll}
\br
\textbf{Systematic effect}   & \textbf{$\Delta\nu/\nu_0\;[10^{-18}]$} & \textbf{$\sigma/\nu_0\;[10^{-18}]$} \\
\mr
Electric quadrupole & 1.7 & 6.2 \\
Phase chirp & 0 & 2.2 \\
Black-body radiation & $-$66.4 & 1.2 \\
Servo offset & 0 & 0.9 \\
Background gas collisions & 0 & 0.6 \\
Quadratic Zeeman (DC) & $-$29.4 & 0.6 \\
AC Stark - probe beam & 2874$^*$ & 0.4 \\
Second-order Doppler & $-$1.3 & 0.4 \\
AC Stark - overshoot & 0 & 0.1 \\
Quadratic Zeeman (AC) & $-$0.3 & 0.1 \\
Trapping RF Stark & $-$0.31 & $0.09$ \\
Trap-induced AC Zeeman & $<0.01$ & $<0.01$ \\
AC Stark - leakage light & $<0.01$ & $<0.01$ \\ 
\mr
\textbf{Total} & $\mathbf{-96.0}$ & $\mathbf{6.9}$ \\
\br
\end{tabular}
\end{indented}
\end{table}
\begin{table*}[t!]
\centering
\caption{Values and uncertainties associated to each ratio that contributes to the absolute frequency calculation of the E3 transition of $^{171}$Yb$^+$ over the period MJD 58664 -- 58674 (July 2019) according to equation~\ref{eq:tai-full}, where $R_0=642\,121\,496\,772\,645.0$.}
\label{tab:2019_tai}
\begin{center}
\resizebox{0.75\textwidth}{!}{
\begin{tabular}{@{}lllll}
\br
\textbf{Ratio}   & \textbf{Contribution} & $\mathbf{r_0}$   & $\mathbf{\left(\frac{r}{r_0}-1\right)\;[10^{-18}]}$ & $\mathbf{u\left(\frac{r}{r_0}-1\right)\;[10^{-18}]}$ \\
\mr
$\frac{f_{\mathrm{Yb}^+;\Delta t_1}}{f_{\mathrm{UTC(NPL)};\Delta t_1}}$ & Ratio at comb & $R_0$ & 2607         & 100                \\
        & Yb$^+$ statistical   &     & --            & 7                  \\
        & Yb$^+$ systematics   &     & 96         & 7                \\
        & Relativistic redshift &  & $-$1187        & 2                  \\ \hline
$\frac{f_{\mathrm{UTC(NPL)};\Delta t_1}}{f_{\mathrm{UTC(NPL)};\Delta t_2}}$ & Maser noise      & 1     & --            & 566                \\
        & Maser drift       &       & $-$588           & 274                 \\ \hline
$\frac{f_{\mathrm{UTC(NPL)};\Delta t_2}}{f_{\mathrm{TAI};\Delta t_2}}$ & UTC(NPL) to TAI  & 1 & 231         & 877                \\ \hline
$\frac{f_{\mathrm{TAI};\Delta t_2}}{f_{\mathrm{TAI};\Delta t_3}}$ & EAL extrapolation  & 1     & --            & 42                 \\ \hline
$\frac{f_{\mathrm{TAI};\Delta t_3}}{f_{\mathrm{SI};\Delta t_3}}$ & TAI to SI second     & 1           & $-$550          & 160                \\
\mr
$\mathbf{\frac{f_{\mathrm{Yb}^+}}{f_{\mathrm{SI}}}}$   & \textbf{Total}        & $\mathbf{R_0}$           & \textbf{610}         & \textbf{1096}                \\
\br
\end{tabular}}
\end{center}
\end{table*}
The optical clock measurement carried out in July 2019 differed from March 2023 mostly in the measurement method of the quadrupole shift, which significantly dominates the systematic uncertainty, as shown in table~\ref{tab:2019_budget}. The quadrupole shift measurement in five magnetic fields was performed against the local $^{87}$Sr lattice optical clock \cite{Hobson2020} acting as a stable reference instead of a second interleaved E2 servo. Moreover, the DC trap electrode configuration generating the electric field gradient had not at the time been optimised for minimal quadrupole shift uncertainty. The difference in electric field gradient is the reason why the shift itself is very different from the other years. Additionally, a correlation was found between the hardware failure of a frequency synthesiser that occurred during the E3 measurement campaign (which resulted in a portion of E3 campaign data being invalidated) and the disagreement within the combined statistical uncertainty of the quadrupole shift measured before and after the campaign. This meant that its final uncertainty had to be inflated to account for this effect.
Another notable difference is the higher uncertainty on the phase chirp shift in 2019: the shift was being compensated in real time, but the level of residual phase shift was not characterised with low enough uncertainty. After performing the more detailed measurements and analysis described in section~\ref{sec:chirp}, it was found that the shift was small enough that it no longer required cancellation in real time, leading to much lower uncertainties in 2022 and 2023. Finally, the probe duration used for the 2019 measurement was 480 ms for both the high and low-power servos.

While in July 2019 the NPL caesium fountain was not operational, the absolute frequency of the E3 transition was still measured via TAI over the period MJD 58664 -- 58674 covering two 5-day reporting periods, during which the optical clock had a 40~\% uptime. The individual ratio contributions are shown in table~\ref{tab:2019_tai}. For this measurement, the frequency comb was referenced directly to the steered output of HM2, the maser generating UTC(NPL) in 2019, whose noise properties are slightly different from those of HM6. Furthermore, this maser was steered once every few days in order to counteract its drift, and these steers were accounted for when calculating the linear drift of the maser. The dominant uncertainty for this measurement comes from the offset of UTC(NPL) from TAI, as it is inversely proportional to the duration of the absolute frequency computation period, which in this case represents only 33~\% of the 1-month TAI evaluation period. The resulting absolute frequency is $642\,121\,496\,772\,645.39(70)$ Hz.

\section{Conclusion}
A full evaluation of the systematic uncertainty budget of the $^2\mathrm{S}_{1/2}\,\rightarrow\, ^2\mathrm{F}_{7/2}$ E3 clock transition in $^{171}$Yb$^+$ has been presented, achieving a relative standard uncertainty of $2.2\times10^{-18}$, currently limited by the uncertainty on the electric quadrupole shift. Moreover, a series of absolute frequency measurements of the E3 transition of $^{171}$Yb$^+$ has been reported, where the lowest fractional uncertainty achieved was $3.7\times~10^{-16}$.

\begin{figure}[h!]
    \centering
    \includegraphics[width=\linewidth]{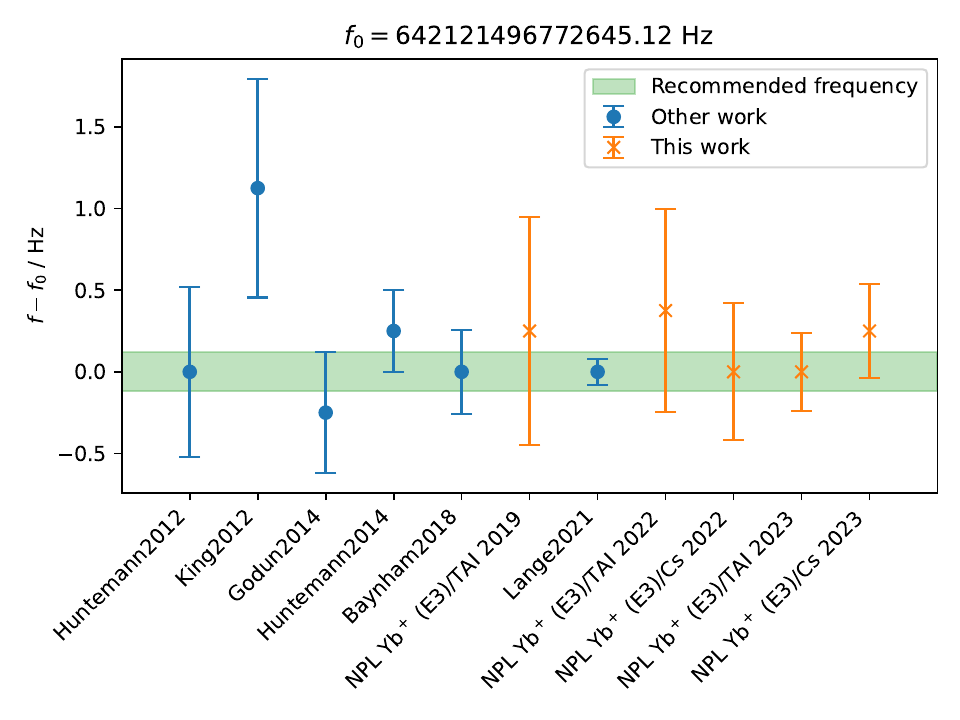}
    \caption{Measurements of the $^{171}$Yb$^+$ E3 transition at NPL during the 2019, 2022 and 2023 campaigns compared to previously published work \cite{huntemann12,King2012,Godun2014,Huntemann14,baynham2018,Lange2021b} and the 2021 recommended value for the frequency. The green band represents the uncertainty on the 2021 BIPM recommended value $f_0$.}
    \label{fig:e3-over-years}
\end{figure}

A graphical summary of all the NPL $^{171}$Yb$^+$ E3 absolute frequency results presented in this paper is displayed in figure~\ref{fig:e3-over-years}, together with other results since 2012 from the literature, showing a very consistent picture in recent years. These new measurements will be submitted for contribution to the next recalculation of the recommended value of the $^{171}$Yb$^+$ E3 absolute frequency performed by the CCL-CCTF Frequency Standards Working Group. Measurements of the $^{171}$Yb$^+$ E3 frequency relative to UTC(NPL) could also be used in the future to contribute to the steering of TAI \cite{Dimarcq2024}.

\ack
The authors would like to thank Sean Mulholland for his careful reading of the manuscript and for helpful technical discussions. Moreover, the authors thank Roxanne Siadat and An Tran for their help in monitoring the frequency comb. This work was supported by the UK government Department for Science, Innovation and Technology through the UK National Measurement System programme.
The authors also acknowledge funding from the European Metrology Programme for Innovation and Research (EMPIR) projects 18SIB05 ROCIT and 20FUN01 TSCAC, co-financed by the
Participating States and from the European Union’s Horizon 2020 research and innovation programme. Support was also received from the Science and Technology Facilities Council as part of the QSNET project (ST/T00598X/1) within the Quantum Technologies for Fundamental Physics Programme. A.T. acknowledges support from the EPSRC Centre for Doctoral Training in Controlled Quantum Dynamics (Grant No. EP/L016524/1).

\section*{References}
\bibliography{References.bib}

\end{document}